\documentclass[aps,amsmath,amssymb,reprint]{revtex4-2}
\usepackage[utf8]{inputenc}
\usepackage{graphicx, float, xcolor, pgf}
\usepackage{physics, siunitx}
\usepackage[caption=false, justification=centerlast]{subfig}

\makeatletter
\let\oldtheequation\theequation
\renewcommand\tagform@[1]{\maketag@@@{\ignorespaces#1\unskip\@@italiccorr}}
\renewcommand\theequation{(\oldtheequation)}
\makeatother

\usepackage{hyperref}
\hypersetup{
  pdfauthor={Amartya Bose},
  pdftitle={Stealthy Disordered Hyperuniform Spin Chains},
  colorlinks=true,
  linkcolor=black,
  citecolor=black,
  urlcolor=black
}

\begin{document}
\title{Quantum Phase Transitions in Long-Range Interacting Hyperuniform Spin Chains in a Transverse Field}
\author{Amartya Bose}
\affiliation{Department of Chemistry, Princeton University, Princeton, New Jersey 08544}
\author{Salvatore Torquato}
\affiliation{Department of Chemistry, Princeton University, Princeton, New Jersey 08544}
\affiliation{Department of Physics, Princeton University, Princeton, New Jersey 08544}
\affiliation{Princeton Institute for the Science and Technology of Materials, Princeton University, Princeton, New Jersey 08544}
\affiliation{Program in Applied and Computational Mathematics, Princeton University, Princeton, New Jersey 08544}

\allowdisplaybreaks

\begin{abstract}
  Hyperuniform states of matter are characterized by anomalous suppression of long-wavelength density fluctuations. While most of the interesting cases of disordered hyperuniformity are provided by complex many-body systems like liquids or amorphous solids, classical spin chains with certain long-range interactions have been shown to demonstrate the same phenomenon. Such systems involving spin chains are ideal models for exploring the effects of quantum mechanics on hyperuniformity. It is well-known that the transverse field Ising model shows a quantum phase transition (QPT) at zero temperature. Under the quantum effects of a transverse magnetic field, classical hyperuniform spin chains are expected to lose their hyperuniformity. High-precision simulations of these cases are complicated because of the presence of highly nontrivial long-range interactions. We perform extensive analysis of these systems using density matrix renormalization group (DMRG) to study the possibilities of phase transitions and the mechanism by which they lose hyperuniformity. Even for a spin chain of length 30, we see discontinuous changes in properties like the ``$\tau$ order metric'' of the ground state, the measure of hyperuniformity and the second cumulant of the total magnetization along the $x$-direction, all suggestive of first-order QPTs. An interesting feature of the phase transitions in these disordered hyperuniform spin chains is that, depending on the parameter values, the presence of transverse magnetic field may remarkably lead to increase in the order of the ground state as measured by the ``$\tau$ order metric,'' even if hyperuniformity is lost. Therefore, it would be possible to design materials to target specific novel quantum behaviors in the presence of a transverse magnetic field. Our numerical investigations suggest that these spin chains can show no more than two QPTs. We further analyze the long-range interacting spin chains via the Jordan-Wigner mapping on to a system of spinless fermions, showing that under the pairwise interacting approximation and a mean-field treatment, there can be at most two quantum phase transitions. Based on these numerical and theoretical explorations, we conjecture that for spin chains with long-range pair interactions that have convergent cosine transforms, there can be a maximum of two quantum phase transitions at zero temperature.
\end{abstract}
\maketitle

\section{Introduction}
The notion of hyperuniformity provides a unifying framework to characterize the large scale structure of systems as disparate as crystals, liquids, and exotic disordered states of matter. A hyperuniform state of matter is one where the density fluctuations at very large length scales are suppressed, and consequently, the structure factor, $\lim_{|\mathbf{k}|\to 0}S(\mathbf{k})=0$. All perfect crystals and perfect quasicrystals are hyperuniform and can be rank ordered in terms of their capacity to suppress large-scale density fluctuations~\cite{Martelli2017,Torquato2003}. Disordered hyperuniform materials are exotic amorphous states of matter that are like crystals in the manner in which their large-scale density fluctuations are anomalously suppressed and yet behave like liquids or glasses in that they are statistically isotropic without any Bragg peaks~\cite{Martelli2017,Torquato2003}.

Classical disordered hyperuniform systems are attracting great attention because they are endowed with novel physical properties~\cite{Hejna2013,Zhang2016,Gkantzounis2017,Froufe-Perez2016,Zito2015,Man2013,Xie2013}. There are far fewer studies of quantum mechanical hyperuniform systems. Some exactly solvable quantum systems, such as free fermion systems~\cite{Feynman1972}, superfluid Helium~\cite{Feynman1956}, the ground state of the fractional quantum Hall effect~\cite{Laughlin1987}, and Weyl-Heisenberg ensembles~\cite{Abreu2017} are hyperuniform~\cite{Torquato2018}. Recently, quantum phase transitions (QPTs) have been studied in nearest-neighbor Ising-chains with hyperuniform couplings~\cite{Crowley2019}. The exploration of the interplay between hyperuniformity, order and quantum fluctuations is a fertile area for research. 

Stealthy hyperuniform systems are ones in which the structure factor is zero for wavenumbers in the vicinity of the origin, i.e., there exists a critical radius $K$, such that, $S(|\mathbf{k}|) = 0$ for all $0<|\mathbf{k}|<K$~\cite{Uche2004a,Batten2008,Torquato2015,Zhang2015}. For sufficiently small $K$, stealthy hyperuniform many-particle systems are the highly degenerate disordered ground states of certain long-ranged interactions~\cite{Torquato2015}. Disordered materials based on such stealthy hyperuniform distribution of particles often possess desirable physical properties, such as photonic band gaps that are comparable in size to photonic crystals~\cite{Florescu2009,Man2013}, transparent dense materials~\cite{Leseur2016}, and optimal transport characteristics~\cite{Froufe-Perez2017,Zhang2016}. The collective coordinate optimization technique has been used to generate disordered stealthy hyperuniform many-particle systems~\cite{Uche2004a,Batten2008,Zhang2015}. Such inverse statistical-mechanical algorithms have also been extended to deal with discrete spin systems~\cite{DiStasio2013}, and to determine long-range Ising interactions that have stealthy hyperuniform classical ground states~\cite{Chertkov2016}. More recently, inverse statistical-mechanical techniques have been applied to quantum mechanical problems to construct the Hamiltonian from the eigenstate~\cite{Chertkov2018}.

Historically spin chains have proven to be fertile grounds for exploration of effects of quantum mechanics. In this paper, we characterize the effects of a transverse field on spin systems whose ground states, in absence of such fields, are stealthily hyperuniform. Quantum fluctuations are typically associated with a loss of order in general. This is true for the usual interactions that decay monotonically with distance. Examples of this include the loss of the ferromagnetic order in Ising model in presence of transverse field and the broadening of the peaks of the pair correlation function $g(r)$ of quantum liquids as compared to their classical counterparts at similar thermodynamic conditions. We demonstrate the very interesting possibility of optimizing magnetic materials such that the quantum effects of an external transverse magnetic field can be used to increase the order of the ground state of the spin system, while still remaining more disordered than the standard antiferromagnetic or ferromagnetic materials. 

In presence of a transverse field, classical mechanics alone cannot account for the physics of a spin chain. Since hyperuniformity is extremely sensitive to the exact nature of the ground state and the interactions involved, it is natural to expect that the transverse field would lead to a loss of hyperuniformity. It is, therefore, interesting to study the quantum mechanism by which hyperuniformity is lost, and the phase transitions involved. In this study, we consider spin chains with long-range interactions that have been optimized to have disordered stealthy hyperuniform ground states in the absence of magnetic field~\cite{Chertkov2016}. We explore these systems using Density Matrix Renormalization Group (DMRG)~\cite{White1992,Schollwock2011,Sirker2005,Jaschke2018,Stoudenmire2012,Schollwock2005,Levin2007,Schollwock2011a,Rotureau2006,Pippan2010} with a particular emphasis on the difference in the physics due to the disordered nature of the classical ground state in absence of the transverse magnetic field and the long-range nature of the interactions. DMRG is one of the best algorithms for dealing with one-dimensional lattice-based problems that are not at the critical points. However, it is typically applied to systems with short-range interactions with open boundary conditions. Recently, DMRG has been applied to study the physics of long-range systems with monotonically decaying interactions with open boundaries~\cite{Saadatmand2018}. Stealthy hyperuniform systems, however, can only be generated in the presence of periodic boundary conditions. Hence, this study involves complex non-monotonic, long-range interactions on periodic boundary conditions, which make the simulations very challenging. All DMRG calculations in this paper were performed using the ITensor library~\cite{ITensor}.

In Sec.~\ref{sec:description}, we describe the system under study, the methods employed and the observables calculated. In Sec.~\ref{sec:results}, we illustrate the most important classes of results obtained through the DMRG simulations. Our results demonstrate that it is possible for the $\tau$ order metric~\cite{Torquato2015, Chertkov2016, Torquato2019} of the ground state to increase with a transverse magnetic field. Our numerical simulations also suggest that these long-range spin systems can sustain no more than two QPTs. Of these, we choose the case with two discontinuous QPTs and illustrate a basis set calculation in appendix~\ref{app:ci_basis}. We carry out further analytical explorations by mapping the system on to a system of spinless fermions via the Jordan-Wigner transformation, which results in a fermionic Hamiltonian with more than pairwise interactions that make the direct solution non-trivial. We analyze the resultant Hamiltonian using a simple approximate model with only the pairwise interaction terms as well as under a mean-field treatment of the terms involving more than pairwise interactions. Consistent with our numerical results, we show in Sec.~\ref{sec:fermion_map} that for both the approximate model and the mean-field Hamiltonian here can be a maximum of two phase transitions. This leads us to a conjecture that for spin chains with long-range pair interactions with convergent cosine transforms, there can be no more than two zero temperature quantum phase transitions. We end this paper with a concluding remarks and outlook for further interesting explorations in Sec.~\ref{sec:conclusions}.

\section{Description of System and Methods Employed}\label{sec:description}
We study one-dimensional (1D) spin systems with long-range interactions. The basic Hamiltonian is evaluated with periodic boundary conditions (PBC) and has the following form on the integer lattice $\mathcal{Z}$: 
\begin{align}
  H &= -\sum_{i}\sum_{1\le r\le R} J_{r} \hat\sigma_z^{(i)} \hat\sigma_z^{(i+r)} + \sum_i -\Gamma \hat\sigma_x^{(i)}.\label{eq:basic_eq}
\end{align}
where $\hat\sigma_{z}^{(i)}$ and $\hat\sigma_{x}^{(i)}$ are the Pauli spin matrices along the z and x directions respectively on the $i$\textsuperscript{th} site, $J_{r}$ is the coupling between two spins separated by $r$ lattice points, and $\Gamma$ is the strength of the transverse field.

We simulate the system for various sets of $J_{r}$ in order to understand how the systems with stealthy hyperuniform ground states in the absence of a transverse field behave with increasing $\Gamma$. These hyperuniform parameters were obtained by~\citet{Chertkov2016} (see Supplementary Material (SM) for the parameters). They are often atypical in the sense that the interaction strength does not necessarily decrease with distance. Therefore, we also show results for simulations where the couplings decay according to inverse power law with the distance.

Due to the increased computational complexity of the DMRG algorithm for periodic systems, all the parameters considered have $N=30$ spins. It is important to note here that in case of simulating long-range interactions in these systems with periodic boundary conditions, one encounters strong finite-size effects. A system with $N=30$ spins might not enough to get rid of these effects. Other investigations have used Ewald summation techniques to take care of this finite-size effect~\cite{Christiansen2019a}. However, here we do not attempt to alleviate this problem in order to maintain consistency with the work on hyperuniform spin chains~\cite{Chertkov2016}.

Because of the long-range interactions, that do not decay with distance and the presence of periodic boundary conditions, the entanglement entropy grows faster than for the regular Ising model in a transverse field. This makes DMRG calculations significantly more difficult to converge and likely to get stuck in other low-lying local minima. Therefore, we perform ten independent simulations of the system. In each simulation, we run twenty DMRG calculations with random initial starting points and taking the state with the minimum energy. We compare results across the various runs and take the lowest energy state as the ground state. Since DMRG is variational in nature, none of the higher energy states can be the true ground state. Performing multiple DMRG calculations often allows us to access other low-lying states. Having an idea of other low-lying states expedites the analysis of the problem using basis set expansions. Because of the multiple DMRG calculations we run, we can converge the wavefunction despite the long-range interactions. However, the convergence gets increasingly difficult in the immediate vicinity of the critical points. This is a well understood limitation of DMRG. At the critical points, the correlation lengths become very large, reducing the effectiveness of the DMRG algorithm. Therefore, these phase transitions and the critical exponents involved cannot be characterized by DMRG. Other methods, such as the Multi-scale Entanglement Renormalization Ansatz (MERA)~\cite{Vidal2007b,Evenbly2009} and the family of Quantum Monte Carlo (QMC) methods, especially the so-called projective QMC methods like diffusion Monte Carlo (DMC), prove to be useful in such studies. Because we are limiting ourselves to chains of length, $N=30$, the locations of the critical points are unlikely to be correct in the thermodynamic limit. However, as our results show, the transitions show a remarkable sharpness, which suggests that these transitions are not artifacts of the finite-size of our systems. Hence our results regarding the possibility of increase in the $\tau$ order metric of the ground state and the variable number of phase transitions would continue to hold qualitatively, even for larger systems.

For each parameter, we first investigate the variation of basic observables like the average energy, and the second cumulant of the transverse magnetization,
\begin{align}
  h_{x} = \frac{1}{N}\left(\expval{M_{x}^{2}} - \expval{M_{x}}^{2}\right)\label{eq:h_x}.
\end{align}
Typically one would use either the average magnetization along the z-direction, $m_{z}$ for the ``standard'' long-range Ising model, or the average magnetization along the x-direction $m_{x}$ defined respectively by
\begin{align}
  m_{z}&=\frac{1}{N}\expval{\sum_{j}\hat\sigma^{(j)}_{z}}\label{eq:avgmag_z}\\
  m_{x}&=\frac{1}{N}\expval{\sum_{j}\hat\sigma^{(i)}_{x}}
\end{align}
However, as we will illustrate, we have found that $h_{x}$ suffers significantly less from finite-size effects than $m_{z}$ for the long-range Ising models, and performs just as well as $m_{x}$ for the hyperuniform cases in identifying the critical points. We also report the structure factor at the origin, the deviation from zero of which is a measure of hyperuniformity,
\begin{align}
  S_0 &\equiv \lim_{k\to 0+} S(k)\\
      &=\frac{1}{N}\left(\expval{M_{z}^{2}} - \expval{M_{z}}^{2}\right)\label{eq:hyperunif}
\end{align}
where $M_{z}$ is the total magnetization along the z-direction, as a function of $\Gamma$. The structure factor at the origin, $S_0 = 0$ for hyperuniform systems. The deviation of $S_0$ from $0$ measures how far the system is from being hyperuniform. As $\Gamma\to\infty$, we should recover a disordered state, irrespective of the exact nature of $J_{r}$ and so, $S_0$ would asymptotically tend to $1$. The degree of order of the ground state is measured using the $\tau$ order metric~\cite{Torquato2015, Chertkov2016, Torquato2019}, defined as follows:
\begin{align}
  \tau &= \sum_{k}\frac{\left(S(k) - S_{\text{ref}}(k)\right)^{2}}{N^{2}}\label{eq:tau},
\end{align}
where $S_{\text{ref}}(k)$ is a reference structure factor. In this paper, we use the structure factor for a Poisson point pattern, $S_{\text{ref}}(k) = 1$ as the reference. The normalization factor of $N^{2}$ is chosen to make most of the results for $\tau$ to be of order unity. With this normalization factor, the antiferromagnetic spin configuration has $\tau\approx 1$, and the ferromagnetic spin configuration has $\tau\approx 2$. The same metric has been used without this normalization factor by~\citet{Chertkov2016} The structure factor of a given ground state is defined as:
\begin{align}
  S(k) &= \frac{1}{N}\sum_{l=1}^{N}\sum_{j=1}^{N}\hat\sigma_{z}^{(l)}\hat\sigma_{z}^{(k)}\exp\left(ik(l-j)\right).
\end{align}

In addition to the hyperuniformity of the ground state, we also want to study the loss of the stealthiness of the hyperuniformity of the ground state. Since, for the examples demonstrated here, the stealthiness extends only to the first non-zero wave-vector, we use the structure factor at the first non-zero wave-vector as a measure of stealthiness in the ground state:
\begin{align}
  S_1 &= S(\Delta k)= S\left(\frac{2\pi}{N}\right).\label{eq:S_1}
\end{align}

Finally, for the spin systems with hyperuniform ground states, we also report the plots of $S(k)$ as a function of $\Gamma$. This allows for a greater clarity in the changes that happen to the ground state before and after the phase transitions.

\section{Results}\label{sec:results}
This section is organized in the following manner. First, we report results for ``standard'' long-range Ising models with interactions that decay with the distance between the spins. These results allow us to set a point of comparison for the hyperuniform spin chains. We illustrate the finite-size effect on the various observables, and demonstrate how the quantum effects of a transverse field reduces the order of the ground state of the system. Thereafter, we examine the systems with interactions optimized to give stealthy hyperuniform ground states. We demonstrate that for systems with these nontrivial long-range interactions, it is possible to generate order from disorder using the quantum effects of a transverse magnetic field. Numerically, we observe no more than two QPTs. To further theoretically explore of the nature of these phase-transitions, we map the spin systems onto chains of spinless fermions. The long-range spin-spin interaction manifests not only in long-distance pairwise interaction terms in the Hamiltonian, but results also in terms involving higher-order non-pairwise interaction terms (interactions involving triplets of spins, quadruplets of spins, and so on). We analyze the resultant Hamiltonian under a very simple long-range pairwise interaction approximation, and a subsequent mean-field treatment of non-pairwise interaction terms, showing that in both cases, a maximum of two phase transitions is possible. This is consistent with our numerical results, and leads us to conjecture that for Ising models with long-range interactions that have convergent cosine transforms, there can be a maximum of two zero-temperature quantum phase transitions.

\subsection{Interactions that Decay with Distance}\label{sec:power_law}
\begin{figure*}
  \subfloat[$m_z$]{\includegraphics{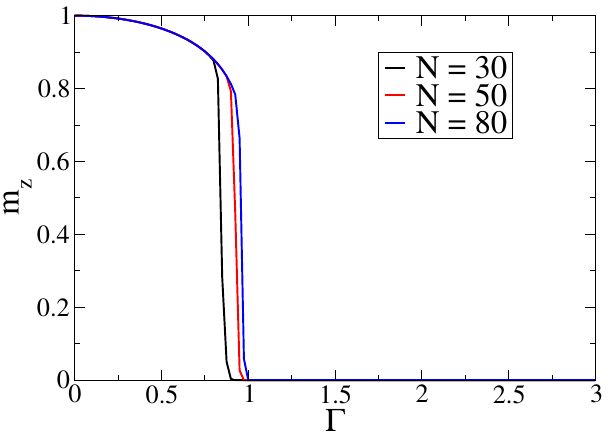}}\hspace{0.1\textwidth}
  \subfloat[$\tau$ order metric]{\includegraphics{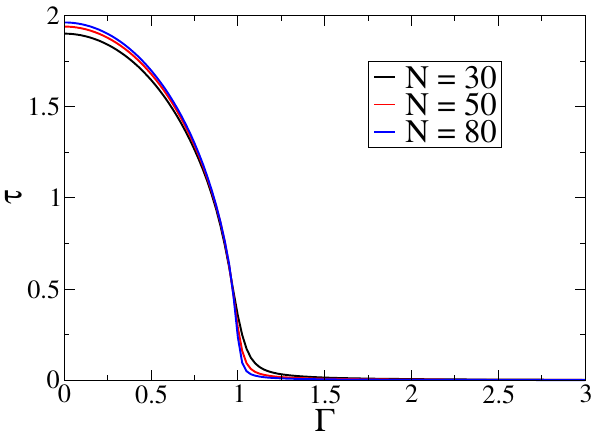}}

  \subfloat[$h_x$]{\includegraphics{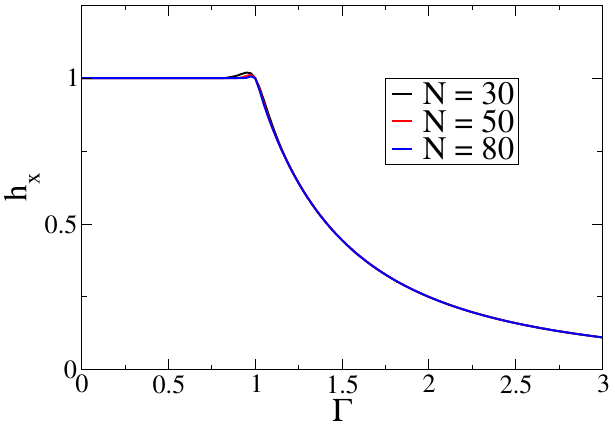}}\hspace{0.1\textwidth}
  \subfloat[$S_0$]{\includegraphics{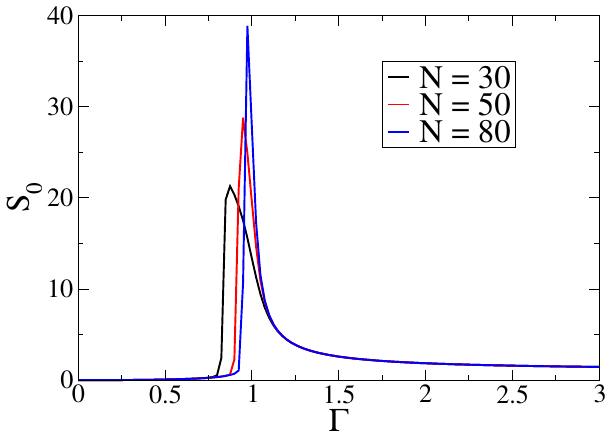}}
  \caption{Size dependence of the critical point using various observables as the order parameters for the ferromagnetic Ising model: (a) Average magnetization along $z$-axis, Eq.~\ref{eq:avgmag_z}; (b) $\tau$ order metric, Eq.~\ref{eq:tau}; (c) Second cumulant of the transverse magnetization, $h_{x}$, Eq.~\ref{eq:h_x}; (d) Measure of hyperuniformity, $S_0$, Eq.~\ref{eq:hyperunif}. The variation of the critical value of $\Gamma$ with $N$ is the least for $h_{x}$. The value of $h_x$ undergoes a maximum for small $N$, which changes to a point of non-differentiability as $N$ becomes larger.}\label{fig:size_dep}
\end{figure*}

\begin{figure*}
  \subfloat[Energy per site]{\includegraphics{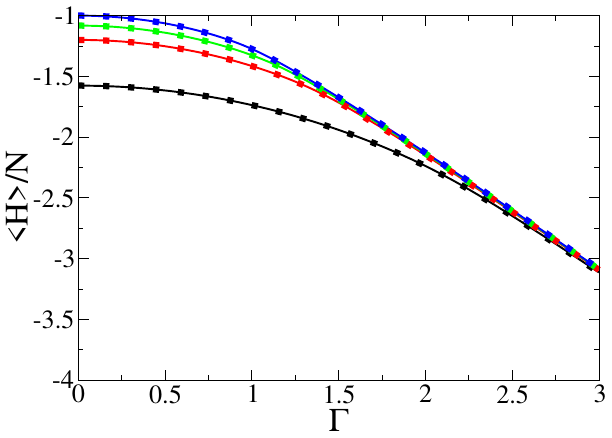}}\hspace{0.1\textwidth}
  \subfloat[$\tau$ order metric]{\includegraphics{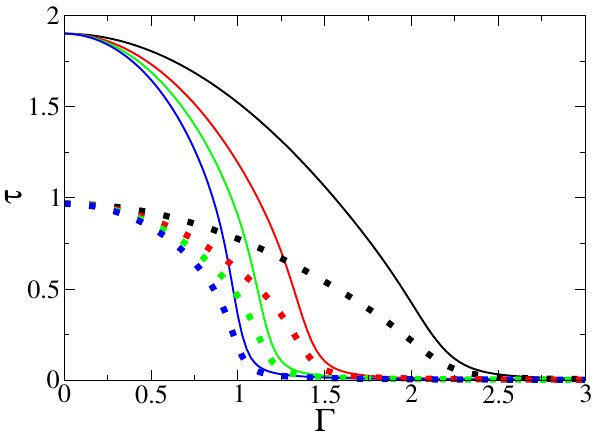}}
  
  \subfloat[$h_x$]{\includegraphics{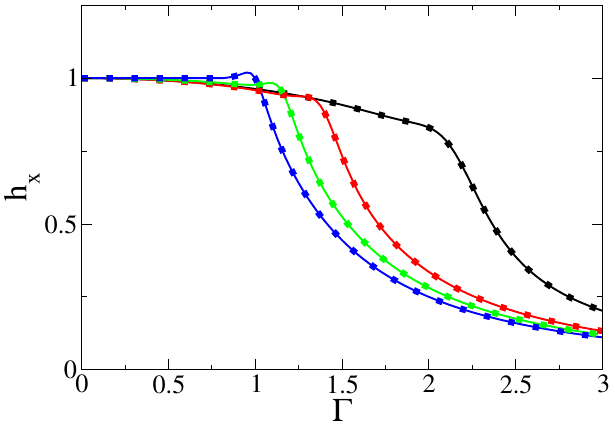}}\hspace{0.1\textwidth}
  \subfloat[$S_0$]{\includegraphics{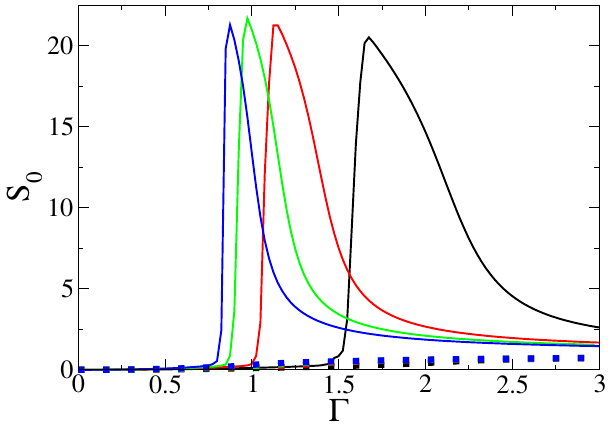}}
  \caption{Basic observables as functions of $\Gamma$: (a) Energy per site; (b) $\tau$ order metric; (c) Second cumulant of the transverse magnetization, $h_{x}$; (d) Degree of hyperuniformity, $S_0$. Black line: $J_{r} = r^{-2}$. Red line: $J_{r} = r^{-3}$. Green line: $J_{r} = r^{-4}$. Blue line: ferromagnetic Ising model. Black markers: $J_{r} = (-1)^{r} r^{-2}$. Red markers: $J_{r} = (-1)^{r} r^{-3}$. Green markers: $J_{r} = (-1)^{r} r^{-4}$. Blue markers: antiferromagnetic Ising model. The degree of hyperuniformity, $S_0$ undergoes a sudden jump for the ferromagnetic cases at the critical point. This jump in $S_0$ is absent in the antiferromagnetic cases.}\label{fig:power_x}
\end{figure*}

For the purposes of comparison to parameters that were specifically optimized to have stealthy hyperuniform ground states, we consider systems with interactions that decay with distance. These models have been extensively studied using various analytic and numerical techniques~\cite{Paulos2016,Hiley1965}. Qualitatively, as decay of the interaction becomes faster, we expect the system to asymptotically approach the nearest neighbor interaction limit, that is the standard Ising model. So, we would expect these systems to undergo a continuous phase transition like the standard Ising model. It is trivial to show using Eq.~\ref{eq:hyperunif} that all of these long-range Ising models have ordered hyperuniform ground state at $\Gamma=0$, owing to the ground states being direct products of eigenvectors of $\hat\sigma_{z}$. 

First, we consider the family of $J_{r} = r^{-a}$ for $a\in\{2,3,4\}$ and $J_{r} = (-1)^{r}r^{-a}$ up to a cutoff radius, $R = 14$ with $N=30$. As a point of comparison, we also include the simulation result for the standard ferromagnetic and antiferromagnetic Ising model. These sets of parameters lead to a ferromagnetic and antiferromagnetic ground states respectively for low values of $\Gamma$. In Fig.~\ref{fig:size_dep}, we show the effect of finite-size on some observables. The average magnetization along the z-axis, $m_{z}$ happens to be a very convenient observable to study the ferromagnetic systems. However, $m_{z}$ suffers from finite-size effect. We note that $h_{x}$ can also be used as a order parameter. This is a measure that is unity for all cases where the ground state is a direct product of eigenstates of the $\hat\sigma_{z}$ operator, that is, when in the ground state, all the spins point either ``up'' or ``down.'' These ``direct-product'' states can either be ``ordered,'' that is either ferromagnetic or antiferromagnetic, or ``disordered'' as we shall see in Sec.~\ref{sec:hyper}. However as we increase $\Gamma\to\infty$, $h_{x}$ decays to zero. It is seen that the second cumulant of the transverse magnetization, $h_{x}$, does not suffer as badly from the finite-size effect. The most notable change in $h_x$ as a function of the system size is that the maximum near the ``critical point'' for small systems seems to change into a point of non-differentiability as the system size gets larger, resulting in a slight movement of the critical point towards the infinite-size limit. The critical points as demonstrated by these different observables converge to the correct thermodynamic value of the critical points in the limit of $N\to\infty$.

In Fig.~\ref{fig:power_x}, we report our simulation results that describe the phase transition in the various ferromagnetic and antiferromagnetic generalized Ising models. The ordered character of the ground state at $\Gamma = 0$ changes at the critical value of the transverse field. The critical point for both the ferromagnetic and antiferromagnetic models happen at exactly the same point. The $\tau$ order parameter decreases monotonically to zero and does not show a discontinuous change at the phase transition. The antiferromagnetic ground state is, of course, less ordered than the ferromagnetic ground state, and this is reflected in the $\tau$ order metric. The degree of hyperuniformity $S_0$ is an order metric for the ferromagnetic chains but not for the phase transition in the antiferromagnetic chains. This is reflected in the sudden, sharp rise in $S_0$ for the ferromagnetic chains shown in Fig.~\ref{fig:power_x}. In all of the inverse power-law interaction Hamiltonians, we see a transition from an ordered hyperuniform state to a disordered non-hyperuniform state. The degree of hyperuniformity, $S_0$, like $m_{z}$, shows a large finite-size effect. Therefore, the critical point should be obtained from the $h_{x}$ curve. Moreover, note that the behavior of all the ferromagnetic and the corresponding antiferromagnetic models in terms of $h_{x}$ is identical, showing that it is a valid order parameter in both cases.

\subsection{Interactions Optimized for Hyperuniformity}\label{sec:hyper}
The interactions that are optimized for hyperuniformity have disordered stealthy hyperuniform ground states at $\Gamma=0$. These ground states, unlike the ones for the typical ferromagnetic long-range interactions that decay with distance, have an average magnetization $m_{z} = 0$ throughout the range of $\Gamma$ and unlike the typical antiferromagnetic long-range interactions, are not ordered. They undergo phase transitions between various ``disordered'' phases with $m_{z} = 0$. The stealthy hyperuniform ground state at $\Gamma=0$, which is a disordered direct-product state, is to be contrasted with the disordered state that is the ground state as $\Gamma\to\infty$. Now, the ground state is a direct product of eigenstates of $\hat\sigma_{x}$ operators. This state, also, has zero total magnetization in the z-direction, but the individual spins are not pointing along the z-direction. We call this the ``disordered quantum'' state.

The spin systems, that we simulated, can be grouped into two broad classes: there are systems with one, or two first-order quantum phase transitions. Of course we would need to increase the system size to truly characterize the phase transitions. However, the sharpness of the discontinuities observed even in the finite sized systems seem to strongly suggest the existence of first-order phase transitions. In the following subsections, we give representative examples of each class. The behavior of stealthiness as measured by $S_1$ is universal across the three classes. Stealthiness is a more sensitive property than hyperuniformity. We will show, in the following sections, that $S_1$ increases faster than $S_0$ but the basic features of $S_1$ are identical to those of $S_0$ in all the classes.

\subsubsection{Parameters with one weak First-Order QPT: No Order from Disorder}
\begin{figure*}
  \subfloat[Energy per site]{\includegraphics{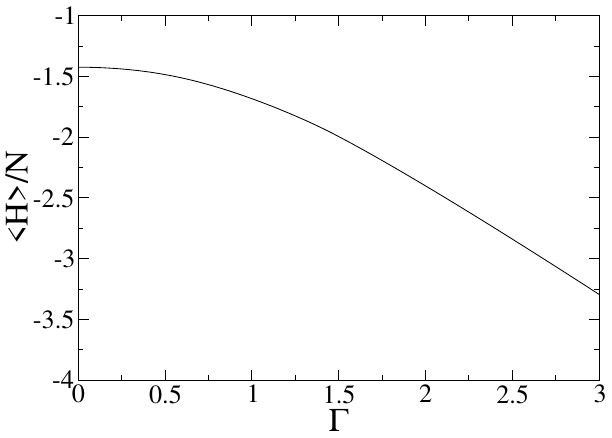}}\hspace{0.1\textwidth}
  \subfloat[$\tau$ order metric]{\includegraphics{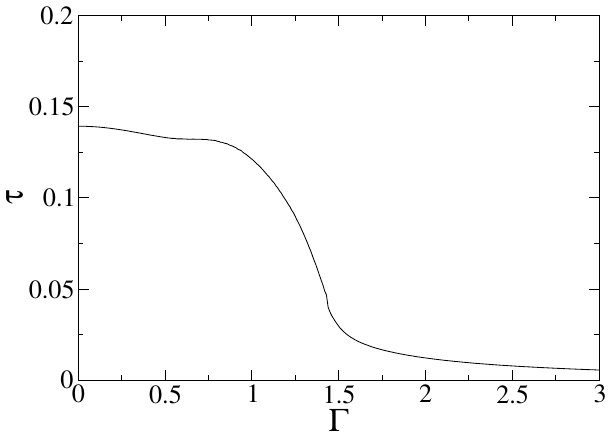}}
  
  \subfloat[$h_x$]{\includegraphics{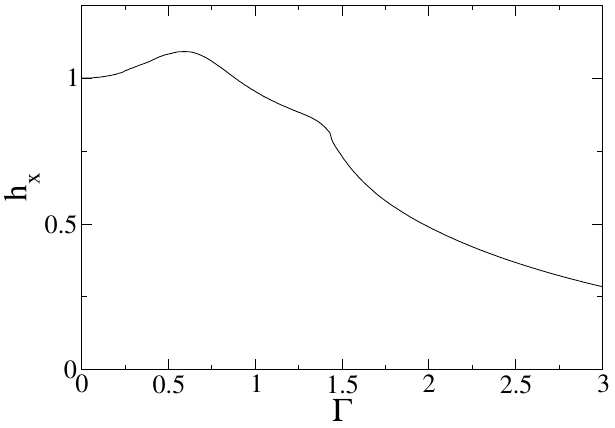}}\hspace{0.1\textwidth}
  \subfloat[$S_0$]{\includegraphics{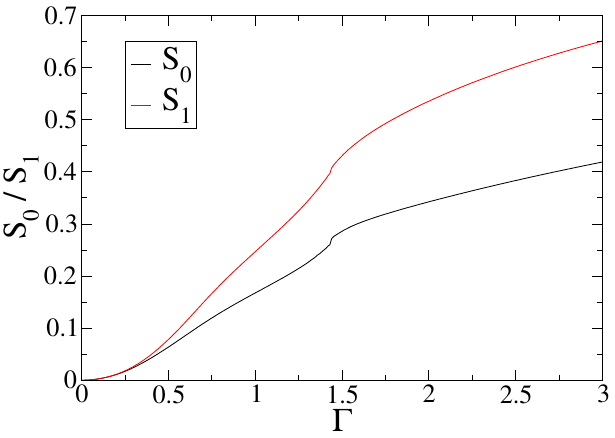}}
  \caption{Basic observables as functions of $\Gamma$: (a) Energy per site; (b) $\tau$ order metric; (c) Second cumulant of the transverse magnetization, $h_{x}$; (d) Degree of hyperuniformity, $S_0$ and $S_1$, Eq.~\ref{eq:S_1}. Parameter demonstrates possibility of a weak first-order QPT.}\label{fig:obs_zero_qpt}
\end{figure*}
\begin{figure}
  \vspace{-0.5cm}
  \includegraphics[scale=0.3]{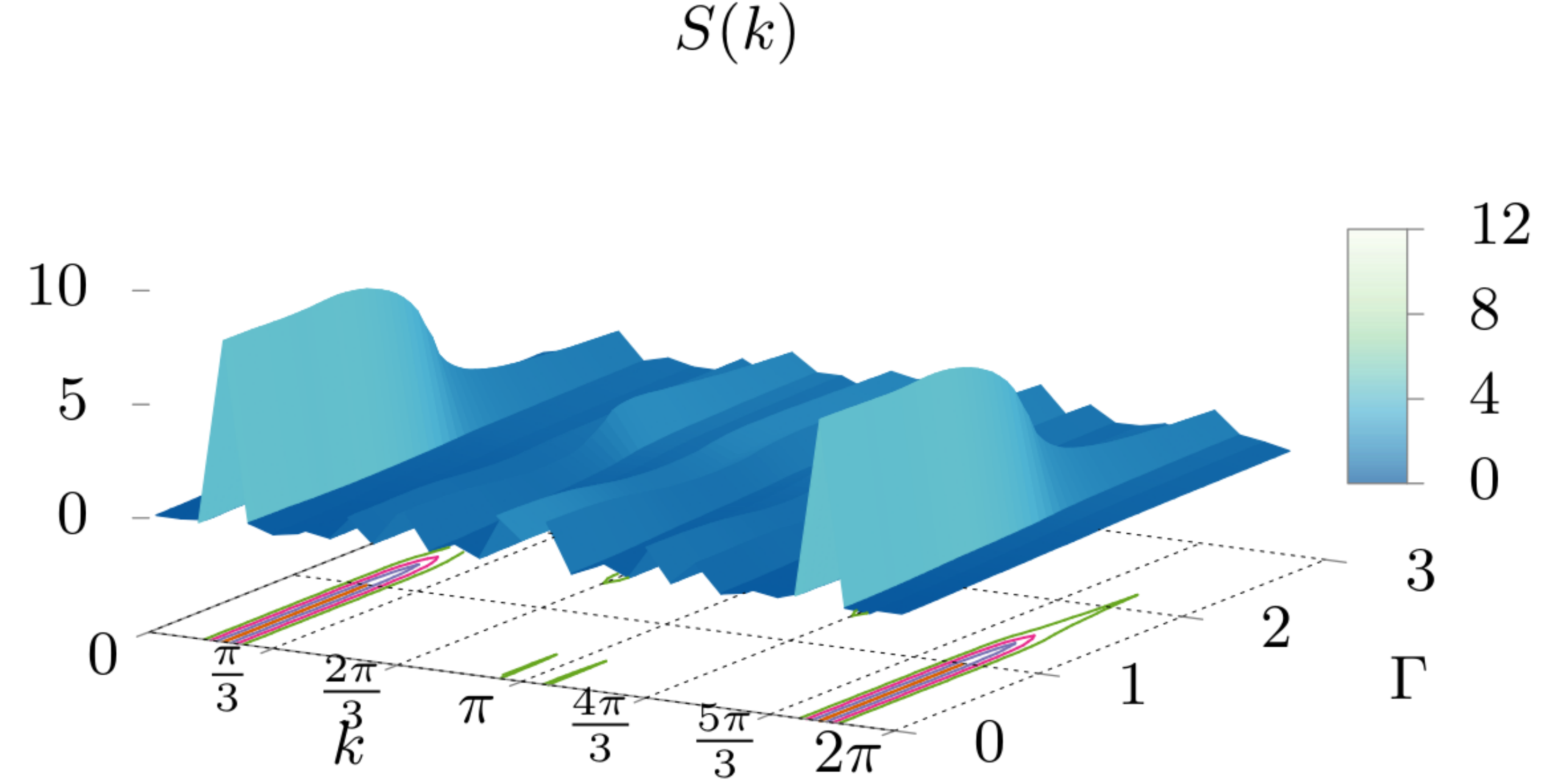}
  \caption{Ground state structure factor as a function of the wave-vector, $k$ and the strength of the transverse field $\Gamma$.}
  \label{fig:Sk_zero_qpt}
\end{figure}
As a first example of the parameters that were optimized for hyperuniformity, consider a system with a weak first-order quantum phase transition (see $J_1(r)$ in SM), where the qualitative features of the $\tau$ order metric are similar to that in the standard long-range Ising model discussed in Sec.~\ref{sec:power_law}. \autoref{fig:obs_zero_qpt} shows the variation of various basic observables as a function of the transverse field. There is a transition at $\Gamma\approx 1.5$ between the classically disordered ground state at ``low'' values of $\Gamma$ and the disordered quantum state at ``large'' values of $\Gamma$.

Notice that the $\tau$ order metric in Fig.~\ref{fig:obs_zero_qpt}(b) is smooth and monotonically decreasing with the transverse field $\Gamma$, similar to the behavior depicted in Fig.~\ref{fig:power_x}(b). This is surprising because the interactions in case of Fig.~\ref{fig:power_x} decay rapidly with distance, whereas the ones in Fig.~\ref{fig:obs_zero_qpt} have been optimized to produce stealthy hyperuniform ground states by long-range interactions. Of course, in the limit of $\Gamma\to\infty$, all structure is lost and the ground state of the Hamiltonian is a direct product of the ground state of the local $\hat\sigma_{x}$ operator. This indicates that the system starts at a disordered hyperuniform ground state at $\Gamma=0$ and continues to lose that order as well. We see that $S_0$ shows a monotonic increase with $\Gamma$, implying that hyperuniformity is degraded. There is a very small ``jump'' in the measures of hyperuniformity, $S_0$ and $S_1$, around $\Gamma\approx 1.5$, which seems to suggest the presence of a weak first-order QPT. $S(k)$ as a function of $\Gamma$ is shown in Fig.~\ref{fig:Sk_zero_qpt}, which, however, seems to show a smooth transition from the disordered classical ground state at low values of $\Gamma$ to the disordered quantum ground state at high values. The smooth decay of the $\tau$ order metric and the small discontinuities in $S_0$ and $S_1$ as functions of $\Gamma$ at the critical point might be a result of the weakness of the first-order QPT in this case.

\subsubsection{Parameters with One First-Order QPT}
\begin{figure*}
  \subfloat[Energy per site]{\includegraphics{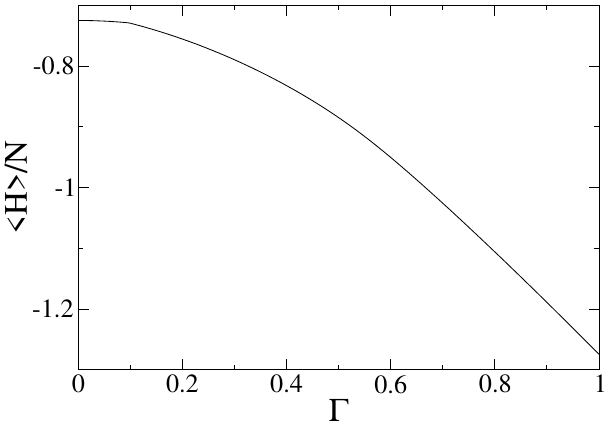}}\hspace{0.1\textwidth}
  \subfloat[$\tau$ order metric]{\includegraphics{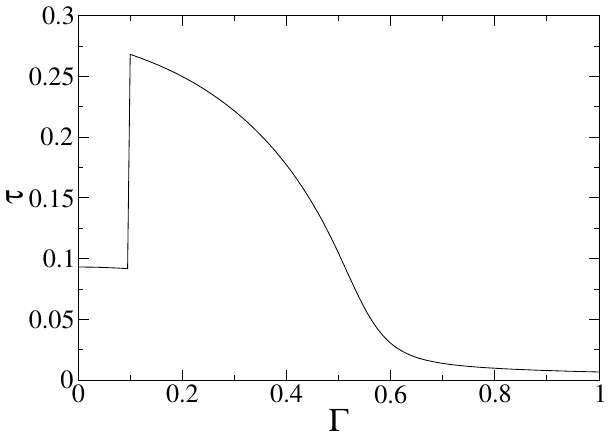}}
  
  \subfloat[$h_x$]{\includegraphics{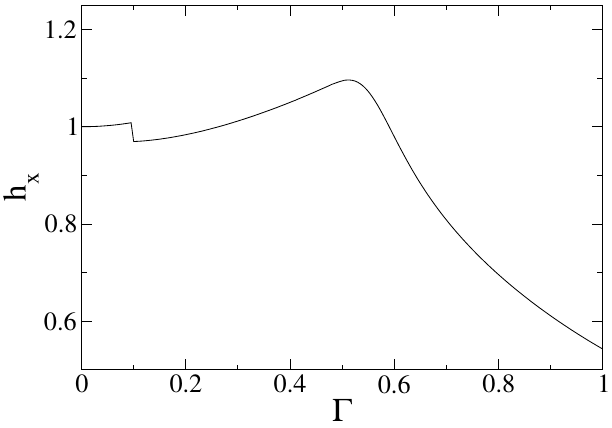}}\hspace{0.1\textwidth}
  \subfloat[$S_0$]{\includegraphics{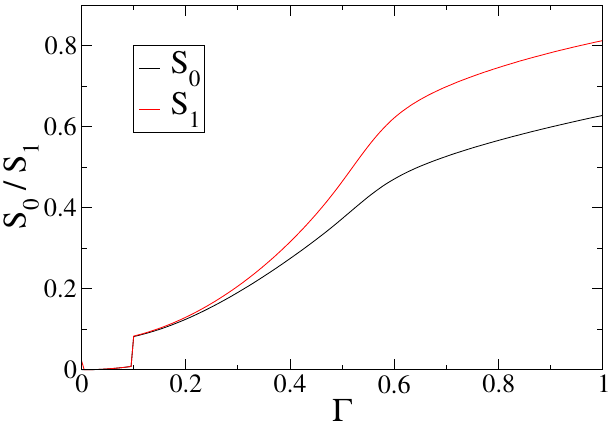}}
  \caption{Basic observables as functions of $\Gamma$: (a) Energy per site; (b) $\tau$ order metric; (c) Second cumulant of the transverse magnetization, $h_{x}$; (d) Degree of hyperuniformity, $S_0$ and $S_1$. Parameter demonstrates possibility of a first-order QPT indicated by sharp discontinuities in (b), (c), and (d) around $\Gamma\approx 0.1$.}\label{fig:obs_one_qpt}
\end{figure*}
\begin{figure}
  \vspace{-0.5cm}
  \includegraphics[scale=0.3]{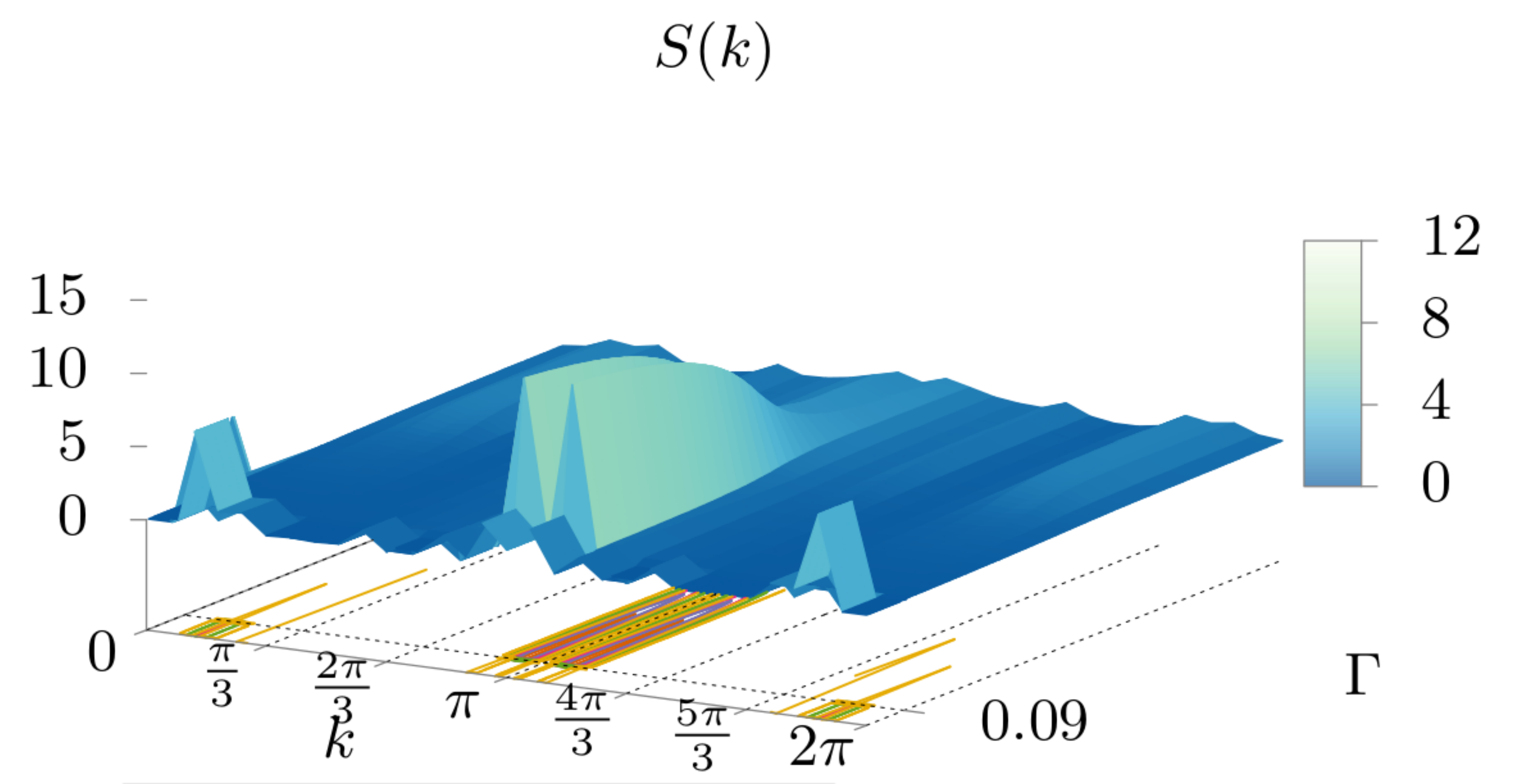}
  \caption{Structure factor as a function of the wave-vector, $k$ and the strength of the transverse field, $\Gamma$. We have marked out the critical value of $\Gamma$ where the phase transition occurs. Between $\Gamma=0.095$ and $\Gamma=0.10$, there is a phase change.}
  \label{fig:contour_one_qpt}
\end{figure}
Next, we consider the $J_{r}$'s which lead to systems with a single first-order phase transition (see $J_2(r)$ in SM). The observables corresponding to one such parameter is shown in Fig.~\ref{fig:obs_one_qpt}. There is a phase transition at $\Gamma=0.09$. It is interesting to note the behavior of the order metric in the vicinity of the phase transition. In this case, the system starts from a relatively disordered, hyperuniform ground state. As the magnetic field is increased, initially the order metric does not undergo any substantial change. Surprisingly, the new structure after the critical point is much more structured than the one with $\Gamma=0$. This order is then smoothly lost as the magnetic field is increased, leading asymptotically to a completely disordered system. In Fig.~\ref{fig:contour_one_qpt}, we provide a surface plot of $S(k, \Gamma)$ to demonstrate the changes in the structure factor as a function of $\Gamma$.

\subsubsection{Parameters with Two First-Order QPTs}
\begin{figure*}
  \subfloat[Energy per site]{\includegraphics{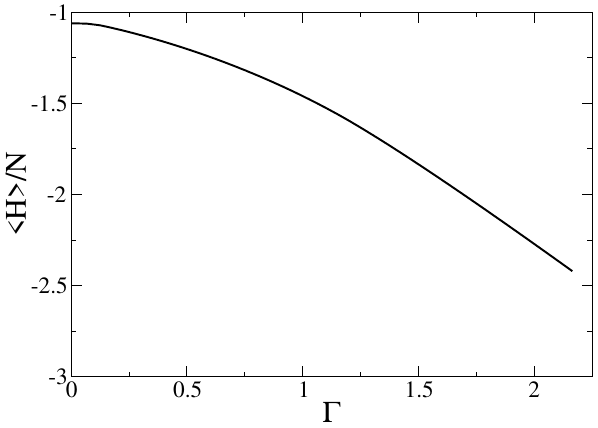}}\hspace{0.1\textwidth}
  \subfloat[$\tau$ order metric]{\includegraphics{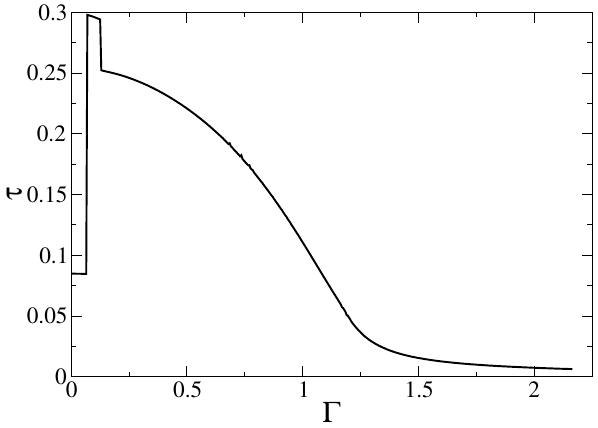}}
  
  \subfloat[$h_x$]{\includegraphics{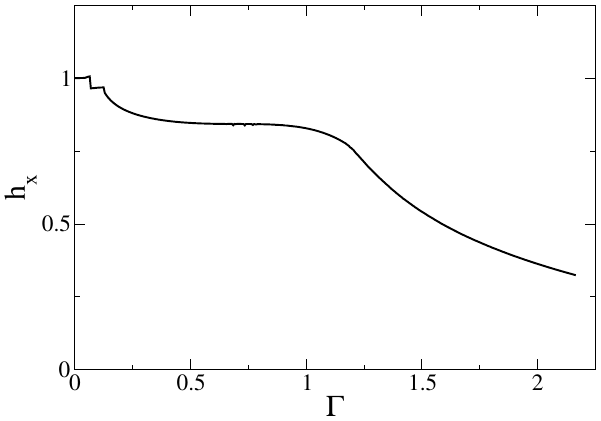}}\hspace{0.1\textwidth}
  \subfloat[$S_0$]{\includegraphics{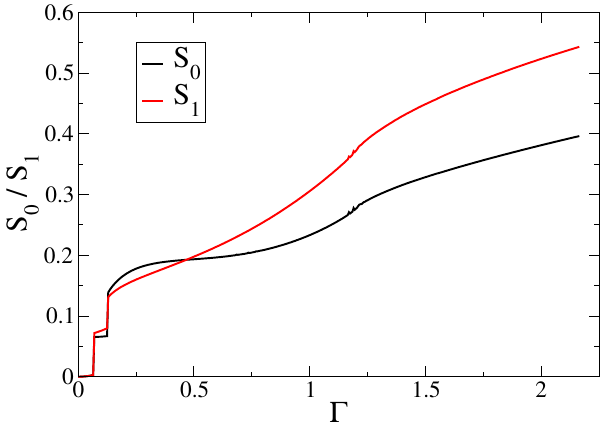}}
  \caption{Basic observables as functions of $\Gamma$: (a) Energy per site; (b) $\tau$ order metric; (c) Second cumulant of the transverse magnetization, $h_{x}$; (d) Degree of hyperuniformity, $S_0$ and $S_1$. Parameter demonstrates possibility of two first-order QPT indicated by sharp discontinuities in (b), (c), and (d) around $\Gamma\approx 0.065$ and $\Gamma\approx 0.125$.}\label{fig:obs_two_qpt}
\end{figure*}

\begin{figure*}
  \subfloat[]{\includegraphics{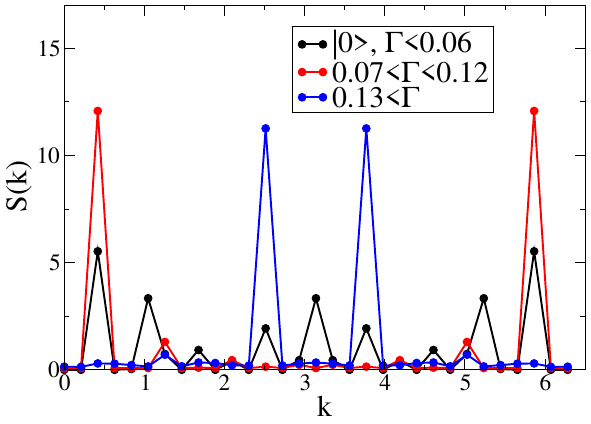}}\hspace{0.1\textwidth}
  \subfloat[]{\includegraphics{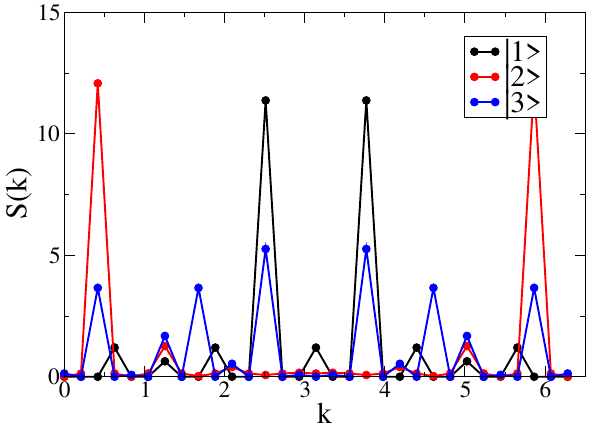}}
  \caption{Structure factors. Left: Ground states for select intervals of $\Gamma$. Ground state before the first critical point $(\Gamma<0.06)$ is labeled $\ket{0}$. Right: Other low-lying states at $\Gamma=0$, labeled $\ket{1}$, $\ket{2}$, and $\ket{3}$ respectively.}
  \label{fig:sk0_low_two_qpt}
\end{figure*}

Finally, there are cases with two first-order phase transitions as demonstrated by the observables in Fig.~\ref{fig:obs_two_qpt} (see $J_3(r)$ in SM). From the $S_0$ plot, it is clear that there are two phase transitions: one between $\Gamma = 0.0675$ and $\Gamma = 0.07$, and another between $\Gamma=0.125$ and $\Gamma=0.1275$. We further analyze this particular parameter using configuration interaction (CI) like basis set expansions to get a better intuition regarding the phase transition. In the ten independent simulations that we performed, there were occasions when the DMRG procedure produced energetically low-lying states which were not the ground state. In Fig.~\ref{fig:sk0_low_two_qpt}, we plot the structure factors of the ground state and the other low-lying states that we encountered. These low-lying states are not relevant to the current discussion, but are useful as additional references for the configuration-integral (CI) like basis set analysis of the phase transitions presented in appendix \ref{app:ci_basis}. As shown in Fig.~\ref{fig:sk0_low_two_qpt}, the ground state goes from $\ket{0}$ before the first phase transition $(\Gamma<0.06)$, to a structure like $\ket{2}$ between the two phase transitions $(0.07 < \Gamma < 0.12)$. Finally, after the phase transition at $\Gamma\approx 1.2$, the structure becomes significantly more similar to $\ket{1}$.
\begin{figure}
  \vspace{-0.5cm}
  \includegraphics[scale=0.3]{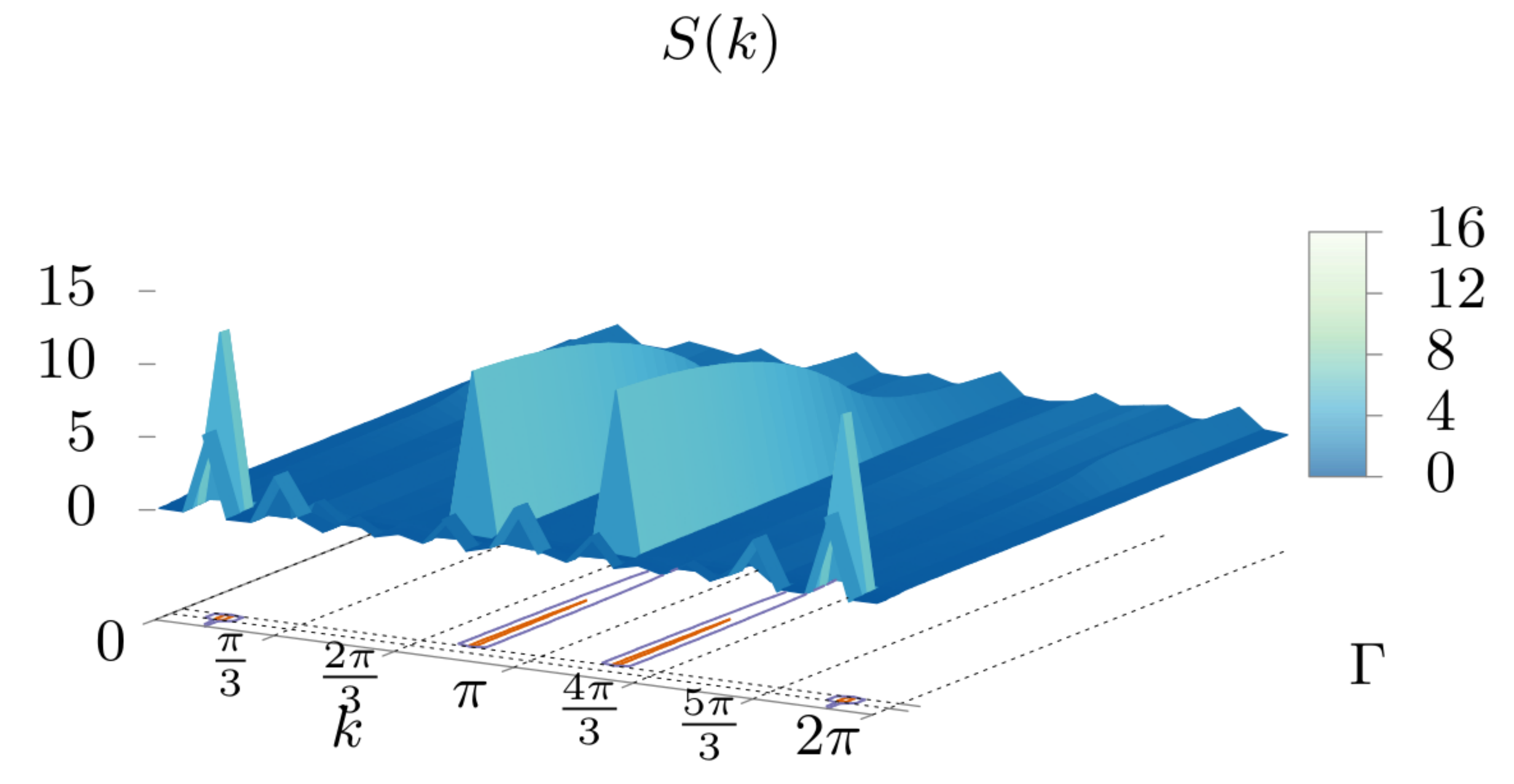}
  \caption{Structure factor as a function of the wave-vector, $k$ and the strength of the transverse field, $\Gamma$. We have marked out critical values of $\Gamma$ where the phase transitions occur. There are two phase transitions, one at $\Gamma\approx 0.07$ and $\Gamma\approx 0.125$, after which there is a steady continuous loss of order.} 
  \label{fig:contour_phasetrans}
\end{figure}

These last two cases are extremely interesting from the fundamental perspective of understanding the impact of a transverse field to these complicated long range interacting spin chains. It is curious that depending on the parameters of the system, phase transitions caused by transverse magnetic fields can serve to increase the order of the ground state as measured by the $\tau$ order parameter. Additionally, the order does not have to necessarily increase. Parameters can be defined where the phase transitions involved have predefined effect on the order metric. As in this case, the system starts off from a disordered hyperuniform state similar to the previous two cases. At the first QPT, there is a sudden increase in order, which decays smoothly till we encounter the second phase transition. Then there is a sudden decrease in order from where the $\tau$ order parameter relaxes smoothly to the disordered state as $\Gamma\to\infty$.

We have numerically demonstrated that there are no cases with more than two QPTs. To provide some analytical motivation for this conjecture, in the next section we analyze the problem using the Jordan-Wigner transform and prove the conjecture for a very simple approximation of the model and the mean-field Hamiltonian.  

\subsection{Analysis using the Jordan-Wigner Mapping}
\label{sec:fermion_map}
Our goal, here, is to analytically diagonalize the Hamiltonian and study the nature of the ground state under a couple of limiting cases. Using a unitary rotation, the Hamiltonian defined in Eq.~\ref{eq:basic_eq}, can be written as:
\begin{align}
  H &= -\sum_{i}\sum_{1\le r\le R} J_{r} \hat\sigma_x^{(i)} \hat\sigma_x^{(i+r)} + \sum_i -\Gamma \hat\sigma_z^{(i)}\label{eq:Htotal_rot}
\end{align}
The Pauli matrices on the same site anticommute and the ones on different site commute. This makes it difficult to apply analytic tools to treat this system. A common method of overcoming this difficulty is to apply a Jordan-Wigner string mapping\cite{Jordan1928} to convert the spin operators to fermionic creating and annihilation operators. The mapping can be summarized as: 
\begin{align}
  \hat\sigma_{z}^{(j)} &= 1 - 2c^\dag_j c_j\\
  \hat\sigma_{+}^{(j)} &= \exp\left(i\pi\sum_{l<j} n_l\right)c_j = \prod_{l<j} \left( 1 - 2c^\dag_l c_l\right)c_j\\
  \hat\sigma_{-}^{(j)} &= \exp\left(-i\pi\sum_{l<j} n_l\right) c^\dag_j = \prod_{l<j} \left( 1 - 2c^\dag_l c_l\right) c^\dag_j
\end{align}
Under this mapping the Hamiltonian, Eq.~\ref{eq:Htotal_rot} would get transformed as follows:
\begin{align}
  H &= -\sum_j\sum_{1\le r\le R} J_r \left(\hat\sigma^{(j)}_+ + \hat\sigma^{(j)}_-\right) \left(\hat\sigma^{(j+r)}_{+} + \hat\sigma^{(j+r)}_{-}\right)\nonumber\\
    &- \Gamma\sum_j \left(1 - 2c^\dag_j c_j\right)\nonumber\\
    &= -\sum_j\sum_{1\le r\le R} J_r \prod^{l<j+r}_{l=j+1}(1-2c^\dag_lc_l)\left( c^\dag_j c^\dag_{j+r} + c^\dag_j c_{j+r}\right.\nonumber\\
    &\left.+ c^\dag_{j+r} c_j + c_{j+r} c_j \right) - \Gamma\sum_j \left(1 - 2c^\dag_j c_j\right)\label{eq:fermH}
\end{align}

Here $c$ and $c^{\dag}$ are the fermionic annihilation and creation operators. The exponentials for $\hat\sigma_{\pm}$ in the Hamiltonian lead to more than pairwise interactions between the fermions. To simplify analysis, we make two approximations: (i) consider only the pairwise interaction terms, and (ii) treat the non-pairwise interaction terms in a mean-field manner.

\subsubsection{Pairwise Interacting Approximation}\label{sec:pairwise}
We begin the analysis by making the crudest simplifying assumption: we consider only the pairwise interaction terms in the Hamiltonian, and make an approximate long-range fermionic model that can be directly solved because it is quadratic: 
\begin{align}
  H &= -\sum_j\sum_{1\le r\le R} J_r \left( c^\dag_j c^\dag_{j+r} + c^\dag_j c_{j+r} + c^\dag_{j+r} c_j + c_{j+r} c_j \right)\nonumber\\
    &- \Gamma\sum_j \left(1 - 2c^\dag_j c_j\right).\label{eq:fermi_2body}
\end{align}

Now, transforming the creation and annihilation operators into Fourier space using $c_j = \frac{1}{\sqrt{N}}\sum_k s_k e^{i k j a}$ and $c^\dag_j = \frac{1}{\sqrt{N}}\sum_k s^\dag_k e^{-i k j a}$, where $a$ is the lattice constant, we get 
\begin{widetext}
  \begin{align}
    H &= \sum_{0<k<\pi} \left( 2 \left( s^\dag_k s_k + s^\dag_{-k}s_{-k} \right) \left( \Gamma - \sum_{1\le r \le R} J_r \cos\left( kar \right) \right) + 2 i \left( s^\dag_{-k} s^\dag_k + s_{-k} s_k \right) \left(\sum_{1\le r \le R} J_r \sin\left( kar \right)\right) \right)\nonumber\\
      &- 2 s^\dag_0s_0\left(\Gamma-\sum_{1\le r\le R} J_r\right) - 2 s^\dag_\pi s_\pi\left(\Gamma-\sum_{1\le r\le R} J_r\cos(\pi a r)\right) - N\Gamma.\label{eq:FourierHamiltonian}
  \end{align}
\end{widetext}

Here we have already gotten rid of terms with double creation (due to the Fermionic nature of the particles) or double annihilation operators. This Hamiltonian can be diagonalized using a standard Bogoliubov transform~\cite{SachdevQPT2011,Jaschke2017a,Zhang2017a}. First, let solve the problem for the ``inner'' part of the Hamiltonian ($k\ne 0$ and $k\ne\pi$). Let $\alpha_k = \left( \Gamma - \sum_{1\le r \le R} J_r \cos\left( kar \right) \right)$ and $\beta_k = \sum_{1\le r \le R} J_r \sin\left( kar \right)$. 
\begin{align}
  H_k &= \begin{pmatrix}s^\dag_k & s_{-k} & s^\dag_{-k} & s_k\end{pmatrix}\begin{pmatrix}\alpha_k & -i\beta_k & 0 & 0\\i\beta_k & -\alpha_k & 0 & 0\\0 & 0 & \alpha_k & i\beta_k\\0 & 0 & -i\beta_k & -\alpha_k\end{pmatrix}\begin{pmatrix} s_k\\s^\dag_{-k}\\s_{-k}\\s^\dag_k\end{pmatrix}
\end{align}

To diagonalize this Hamiltonian, consider the Bogoliubov transform defined by $s_{k} = u_{k}\gamma_{k} + i v_{k} \gamma_{-k}^{\dag}$, where $u_{k} = u_{-k}$, $v_{k} = -v_{-k}$ and $u^{2}_{k} + v^{2}_{k} = 1$. We also define a single parameter, $\theta_{k}$ such that $u_{k} = \cos\left(\frac{\theta_{k}}{2}\right)$ and $v_{k} = \sin\left(\frac{\theta_{k}}{2}\right)$. The solution for all $k\ne 0, \pi$ is given by $\tan\left(\theta_{k}\right) = \frac{\beta_{k}}{\alpha_{k}}$. The eigenvalues of the matrix are given by $\pm\sqrt{\alpha_{k}^{2} + \beta_{k}^{2}}$. For $k=0$ or $k=\pi$, the Hamiltonian in Eq.~\ref{eq:FourierHamiltonian} is already diagonal. Putting everything together, the final diagonalized Hamiltonian is 
\begin{align}
  H &= \sum_{k\ne 0,\pi} 2\sqrt{\alpha_k^2+\beta_k^2}\left(\gamma^\dag_k\gamma_k - \frac{1}{2}\right)\nonumber\\
    &- 2\left( \Gamma - \sum_{1\le r\le R} J_r \right) s^\dag_0s_0\nonumber\\
    &- 2\left( \Gamma - \sum_{1\le r \le R} J_r\cos(\pi a r) \right) s^\dag_\pi s_\pi
\end{align}

Now, all the $k$-modes are independent. Therefore, we can write down the ground state of the Hamiltonian in mixed Bogoliubov (space of $s_k$) / non-Bogoliubov (space of $\gamma_{k}$) space. We only keep $s_{0}$ and $s_{\pi}$ in the standard space and talk about the rest of the modes in terms of $\gamma_{k}$. We define the vacuum $\ket{}$ as a state which is annihilated by $\gamma_{k}$ for all $k\ne 0$ and $k\ne\pi$, and also by $s_{0}$ and $s_{k}$. Because the Hamiltonian is basically just the harmonic oscillator Hamiltonian for all modes with $k\ne 0$ or $k\ne\pi$, the vacuum state is the ground state for the Hamiltonian. The phase transitions in the model are caused by changes in the occupation of the $k=0$ or the $k=\pi$ modes. If $\Gamma>\sum_{1\le r\le R}J_{r}$, then the $k=0$ state being populated lowers the energy, and the ground state must have this state populated. Similarly if $\Gamma>\sum_{1\le r\le R} J_{r}\cos(\pi a r)$, then $k=\pi$ must be populated. This change in the nature of ground state is a phase transition. 

Consider the nearest neighbor ferromagnetic $\left(R = 1, J_{r}=J>0\right)$ and antiferromagnetic $\left(R=1, J_{r}=J<0\right)$ Ising models with the lattice constant $a=1$ as trivial examples of the mapping and cases where the model is exact. For $0\le\Gamma<|J|$, in case of the ferromagnetic Ising model, the $k=\pi$ mode is populated. On the other hand, for the antiferromagnetic Ising model, the $k=0$ mode is populated for $0\le\Gamma\le|J|$. When $\Gamma>|J|$, in case of the ferromagnetic Ising model, the $k=0$ mode gets populated in addition to the $k=\pi$ mode, and in case of the antiferromagnetic case, the $k=\pi$ mode gets populated. This represents a phase transition, with the critical value of $\Gamma=|J|$, that happens by different mechanisms in the two cases: the occupation of the $k=0$ mode in the ferromagnetic case, and the $k=\pi$ mode in the antiferromagnetic case.

Now, we come to the question of the number of such phase transitions. If any one of $\sum_{1\le r\le R} J_{r}$ and $\sum_{1\le r\le R}J_{r}\cos(\pi a r)$ is greater than zero, then there would be one phase transition, and if both of them are greater than zero, then there would be two phase transitions with the critical points being at these values of $\Gamma$. There is no other variable phase transition that is possible because the $k\ne 0$ and $k\ne\pi$ modes are always in the vacuum state. Thus, consistent with the numerical results that we have encountered in the previous sections, this approximate model also allows a maximum of two phase transitions. We would like to emphasize that this condition on $J_{r}$ for the number of phase transitions is only valid for the current approximate model. It is not valid in general for the long-range spin system, for which closed-form analytic solutions do not exist.

Observables can be calculated by mapping them onto the spinless fermions by using the Jordan-Wigner mapping. As an example, we choose $\tfrac{1}{N}\sum_{1\le j\le N} \expval{\sigma_{z}^{(j)}}$. This would be useful in doing the mean-field analysis in the next section. In the following $\gamma_{0} = s_{0}$ and $\gamma_{\pi} = s_{\pi}$. 
\begin{widetext}
  \begin{align}
    \frac{1}{N}\sum_{1\le j\le N} \expval{\sigma_{z}^{(j)}} &= \frac{1}{N}\sum_{k}\left(1 - 2\expval{s^{\dag}_{k}s_{k}}\right) = 1 - \frac{2}{N}\sum_{k}\expval{\left(u_{k}\gamma^{\dag}_{k} - i v_{k}\gamma_{-k}\right)\left(u_{k}\gamma_{k} + i v_{k}\gamma^{\dag}_{-k}\right)}\nonumber\\
                                                            &= 1 - \underbrace{\frac{2}{N}\sum_{k\ne0, \pi} \left(u^{2}_{k}\expval{\gamma^{\dag}_{k}\gamma_{k}} - v^{2}_{k}\expval{\gamma^{\dag}_{-k}\gamma_{-k}}\right)}_{\text{term
                                                            A}} - \frac{2}{N}\sum_{k\ne 0, \pi} v^{2}_{k} - \frac{2}{N}\left(\expval{\gamma^{\dag}_{0}\gamma_{0}} + \expval{\gamma^{\dag}_{\pi}\gamma_{\pi}}\right)\label{eq:A_zero}\\
                                                            &= 1 -  \frac{2}{N}\sum_{k\ne 0, \pi} v^{2}_{k} - \frac{2}{N}\left(\expval{\gamma^{\dag}_{0}\gamma_{0}} + \expval{\gamma^{\dag}_{\pi}\gamma_{\pi}}\right)\label{eq:v_sq}\\
                                                            &= 1 - \frac{1}{N}\sum_{k\ne 0, \pi}\left(1-\frac{\alpha_{k}}{\sqrt{\alpha^{2}_{k} + \beta^{2}_{k}}}\right) - \frac{2}{N}\left(\expval{\gamma^{\dag}_{0}\gamma_{0}} + \expval{\gamma^{\dag}_{\pi}\gamma_{\pi}}\right)\label{eq:v_sq_solved}\\
                                                            &= \frac{2}{N} + \frac{1}{N}\sum_{k\ne 0, \pi}\frac{\alpha_{k}}{\sqrt{\alpha^{2}_{k} + \beta^{2}_{k}}} - \frac{2}{N}\left(\expval{\gamma^{\dag}_{0}\gamma_{0}} + \expval{\gamma^{\dag}_{\pi}\gamma_{\pi}}\right)\label{eq:v_sq_simplified}
  \end{align}
\end{widetext}

Since no $k$-modes apart from $k=0$ and $k=\pi$ are populated with Bogoliubov fermions, the term marked ``A'' is zero in Eq.~\ref{eq:A_zero}. The equality between Eq.~\ref{eq:v_sq} and Eq.~\ref{eq:v_sq_solved} is obtained by noting that $\tan(\theta_{k})=\frac{\beta_{k}}{\alpha_{k}}$ and $v_{k} = \sin\left(\frac{\theta_{k}}{2}\right)$, while the third equality is a result of having $N-2$ modes with $k\ne 0, \pi$. Similar expressions can be derived for other observables.

\subsubsection{Mean-field Analysis}\label{sec:meanfield}
For the mean-field analysis of Eq.~\ref{eq:fermH}, let us assume that at the $i^{\text{th}}$ iteration, the value of $\frac{1}{N}\sum_{1\le l\le N}\expval{1-2c_{l}^{\dag} c_{l}} = g_{i}$. We will use this assumption to solve the Hamiltonian as a function of $g$:
\begin{align}
  H(g) &= -\sum_j\sum_{1\le r\le R} J_r g^{r-1}\left( c^\dag_j c^\dag_{j+r} + c^\dag_j c_{j+r} \right.\nonumber\\
       &\left.+ c^\dag_{j+r} c_j + c_{j+r} c_j \right) - \Gamma\sum_j \left(1 - 2c^\dag_j c_j\right)\label{eq:SCFH}
\end{align}
Of course, the analysis of Eq.~\ref{eq:SCFH} is simplified through the observation that it is isomorphic with Eq.~\ref{eq:fermi_2body} and consequently, Eq.~\ref{eq:FourierHamiltonian} under the transformation $J_{r} \to \tilde{J}_{r} = J_{r} g^{r-1}$. To start the process, we use the $g_{0}$ obtained under the pairwise interaction approximation, using Eq.~\ref{eq:v_sq_solved}. If we are only interested in the value of $\frac{1}{N}\sum_{1\le l\le N}\expval{1-2c_{l}^{\dag}c_{l}}$, it might be possible to directly search for the root of the following equation:
\begin{align}
  g &= \frac{2}{N} + \frac{1}{N}\sum_{k\ne 0, \pi} \frac{\tilde\alpha}{\sqrt{\tilde\alpha^{2} + \tilde\beta^{2}}}\nonumber\\
    &- \frac{2}{N}\left(\expval{\gamma^{\dag}_{0}\gamma_{0}} + \expval{\gamma^{\dag}_{\pi}\gamma_{\pi}}\right)\\
  \text{where }\tilde\alpha &= \Gamma - \sum_{1\le r\le R} J_{r} g^{r-1} \cos(kar)\\
  \tilde\beta &= \sum_{1\le r\le R} J_{r} g^{r-1} \sin(kar).
\end{align}

However, here we are interested in characterizing the number of phase transitions in the ground state as the transverse field is increased. This can be achieved much more simply by observing that because the Hamiltonian at every step is isomorphic to Eq.~\ref{eq:fermi_2body} at every step of the self-consistent field procedure, the constraint on the maximum number of phase transitions in the model would hold. Thus, even after solving the long range problem, in a self-consistent manner, we expect that there cannot be more than two phase transitions, which is consistent with our numerical exploration.

\section{Conclusions}\label{sec:conclusions}
Long-range Ising models can exhibit very rich physics, especially when the long-range couplings do not decay with distance. Such atypical long-range couplings are essential to ensuring that spin systems have stealthy hyperuniform ground states in the absence of any transverse field. Spin systems with hyperuniform ground states, thus, show very interesting physics that is qualitatively different from standard long-range Ising models with interactions that decay with distance. We have presented here one of the first analyses of the physics of stealthy, hyperuniform spin systems with nontrivial, non-monotonic, long-range interactions under a transverse magnetic field. We demonstrate numerically that for these systems, unlike standard long-range Ising models, the number of phase transitions is not fixed. Their loss of hyperuniformity, in presence of transverse fields, is not always accompanied by a loss of order, as measured by the $\tau$ order metric. This feature is very unusual and leads to the possibility of designing hyperuniform materials whose ground-state in the presence of external transverse magnetic fields can be more ordered than that in absence of external the field. We also showed that the rate of loss of the property of stealthiness is much faster than that of the loss of hyperuniformity, proving that stealthiness is a much more delicate property.

To better theoretically understand the phase transitions we identified numerically, we have analyzed the long-range Ising spin models using the Jordan-Wigner mapping Hamiltonian. Under this mapping, the long-range spin-spin interactions manifest themselves as non-pairwise interaction terms. We showed that for a generic long-range Ising model, under the pairwise interaction and mean-field ``approximations,'' there can be a maximum of two phase transitions, which is consistent with our numerical results. Therefore, we conjecture that a long-range pairwise interacting 1D Ising spin chain with arbitrary couplings for which a discrete cosine transform is convergent, can show at most two quantum phase transitions at zero temperature.

Future work on understanding the critical scaling of these phase transitions using MERA or QMC should lead to further insights into the nature of these long-range hyperuniform spin chains. It would also be interesting to study these systems under a combination of longitudinal and transverse fields to map out the full phase diagram. Such studies would prove useful in the design of novel materials with simple models.

\section*{Acknowledgments}
We thank Roberto Car for many insightful discussions. A.~B. acknowledges support of the Computational Chemical Center: Chemistry in Solution and at Interfaces funded by the U.\,S. Department of Energy under Award No.\,DE-SC0019394. S.\,T. acknowledges the support of the National Science Foundation under Grant No.\,DMR-1714722.

\appendix
\section{Explorations using Configuration Interaction (CI) expansions}\label{app:ci_basis}
\begin{figure*}
  \subfloat[CIS Energy]{\includegraphics{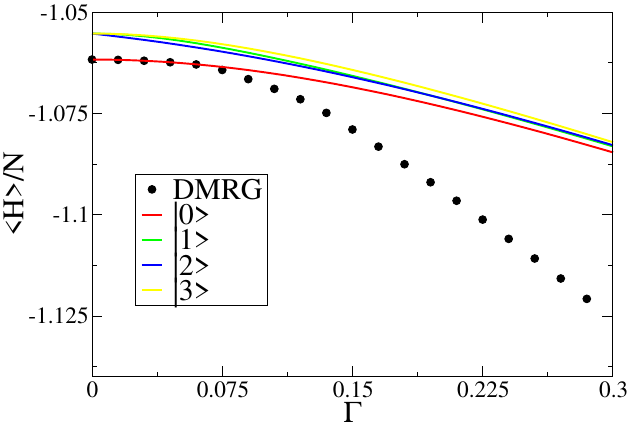}}\hspace{0.1\textwidth}
  \subfloat[CISD Energy]{\includegraphics{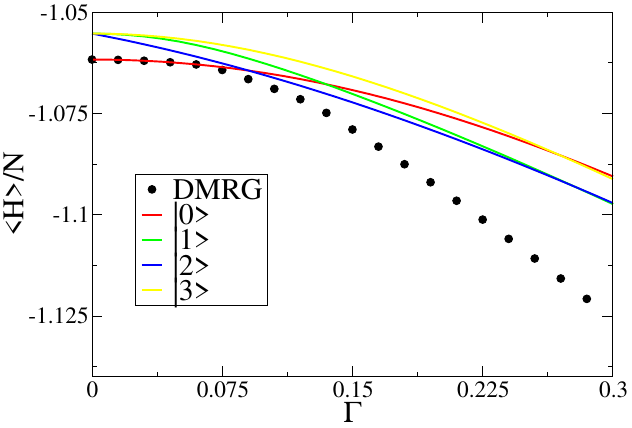}}
  
  \subfloat[CISDT Energy]{\includegraphics{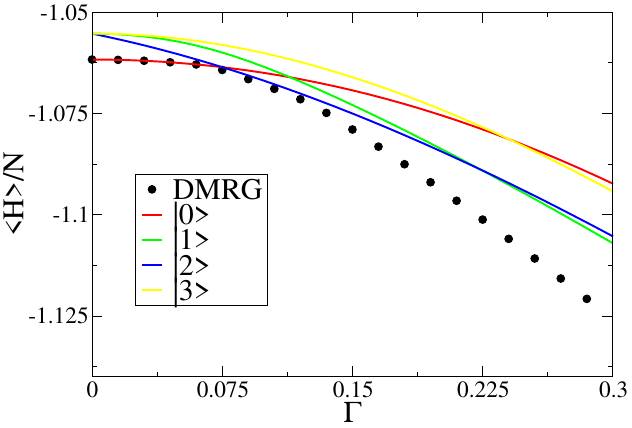}}\hspace{0.1\textwidth}
  \subfloat[CISDTQ Energy]{\includegraphics{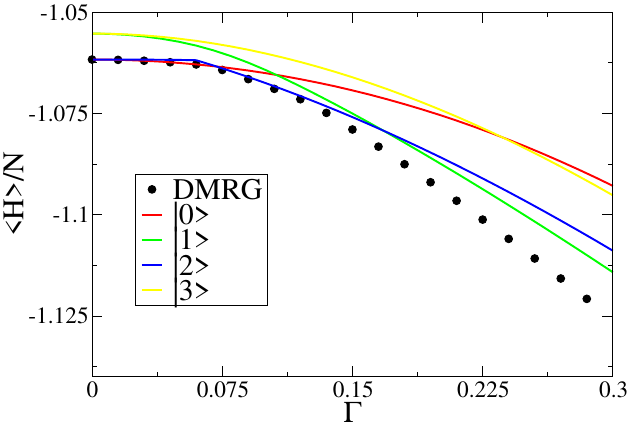}}
  \caption{Energy per site with respect to different references.}\label{fig:ci_trunc}
\end{figure*}

The Hamiltonian can be decomposed in the following manner:
\begin{align}
  H &= \underbrace{-\sum_{i}\sum_{1\le r\le R} J_{r} \hat\sigma_z^{(i)} \hat\sigma_z^{(i+r)}}_{H_{\text{ref}}} + \underbrace{\sum_i -\Gamma \hat\sigma_x^{(i)}}_{V}\label{eq:Htotal}\\
  H_{\text{ref}}|\phi\rangle &= E_0|\phi\rangle
\end{align}
Like in the DMRG simulations, we evaluate the Hamiltonian in a PBC. We have already solved for the ``reference'' $|\phi\rangle$, which is the classical long-range Ising problem with $\Gamma = 0$. Now, we expand the true ground state wave function $|\psi\rangle$ in a configuration integral expansion: 
\begin{align}
  |\psi\rangle &= c_0|\phi\rangle + \sum_{\alpha} c_{\alpha}|\phi_{\alpha}\rangle + \sum_{\alpha,\beta} c_{\alpha,\beta}|\phi_{\alpha,\beta}\rangle + \ldots
\end{align}
where $|\phi_{\alpha}\rangle$ is obtained by flipping the $\alpha^\text{th}$ spin with respect to $|\phi\rangle$, $|\phi_{\alpha,\beta}\rangle$ is obtained by flipping the $\alpha^\text{th}$ and $\beta^\text{th}$ spins, so on.

We solve the eigenvalue equation representing the Hamiltonian in the orthonormal basis described above. First, consider the matrix elements of $H_{\text{ref}}$. Since $H_{\text{ref}}$ is a function of $\hat\sigma_z$ and the basis vectors are eigenvectors of $\hat\sigma_z$, $H_{\text{ref}}$ is diagonal in the basis defined.

Though the basis consisting of all possible excitations is complete, the smaller basis set obtained using a truncated number of excitations is not. As a method of exploring the nature of the ground state qualitatively, we use not just $\ket{0}$ as reference but also $\ket{1}$, $\ket{2}$, and $\ket{3}$. Of course, in the full basis, the results should be independent of the reference used. To gain a better understanding, we truncate the expansion such that, the bases defined on all of the references are non-intersecting. So, the Hamiltonian matrix defined in terms of all the references would have a block diagonal structure, implying that we can solve the problem for each of the references independently. Additionally this allows us to probe into the nature of the ground state. We can now make qualitative statements about which reference the ground state looks like. Note in Fig.~\ref{fig:ci_trunc}, for $ \ket{2}$, the inclusion of quartic excitations includes vectors that are similar to $\ket{0}$. That is why, though up to CISDT, $\ket{2}$ is higher in energy than $\ket{0}$ at $\Gamma = 0$, once the quartic excitations are included, using $\ket{2}$ as reference, we can get the correct behavior in the low $\Gamma$ region as well.

Obviously at $\Gamma$ close to $0.3$ the results are not converged. However we are interested in the nature of the ``crossovers.'' \autoref{fig:ci_trunc} demonstrates that if we consider only single excitations, then there is no phase transition. The ground state remains similar to $\ket{0}$ over the range of $\Gamma$ considered. However as soon as we introduce more excitations, phase transitions start appearing. We notice that for the CISDT calculations, there is a cross-over from a $\ket{0}$-like ground state to a $\ket{2}$-like ground state at around $0.074$, followed by a cross-over from a $\ket{2}$-like ground state to a $\ket{1}$ like ground-state at $0.22$. The $\ket{0}$-like ground-state to $\ket{2}$-like ground-state transition lies within the range of $\Gamma$ where the calculations were better converged. Therefore, we get an estimate of the critical value of $\Gamma$ that is in good agreement with the estimate from DMRG calculations (0.0675 -- 0.07). Though the second phase transition is predicted accurately, the critical value of $\Gamma$ is significantly over-estimated because the expansion calculations are not converged.

\end{document}